\documentclass[twocolumn]{article}
\usepackage{epsfig} 
\usepackage{color} 
\usepackage{graphicx}
\usepackage{amssymb,amsmath}

\def\QED{~\rule[-1pt]{5pt}{5pt}\par\medskip}
\begin{document}

\title{Chaotic Transitions in Wall Following Robots 
\thanks{ The research reported in this document/presentation was performed in connection with contract/instrument W911QX-09-C-0055 with the U.S. Army Research Laboratory.  The views and conclusions contained in this document/presentation are those of the authors and should not be interpreted as presenting the official policies or position, either expressed or implied, of the U.S. Army Research Laboratory or the U.S. Government unless so designated by other authorized documents.  Citation of manufacturer’s or trade names does not constitute an official endorsement or approval of the use thereof.  The U.S. Government is authorized to reproduce and distribute reprints for Government purposes notwithstanding any copyright notation hereon. } }
\author{Harry W. Bullen IV and Priya Ranjan}
\date{ \today}
\maketitle{}

\begin{abstract}
In this paper we examine how simple agents similar to Braitenberg vehicles can exhibit chaotic movement patterns.
The agents are wall following robots as described by Steve Mesburger and Alfred Hubler in their paper ``Chaos in Wall Following Robots''.
These agents uses a simple forward facing distance sensor, with a limited field of view ($\Phi$) for navigation.
An agent drives forward at a constant velocity and uses the sensor to turn right when it is too close to an object and left when it is too far away.

For a flat wall the agent stays a fixed distance from the wall and travels along it, regardless of the sensor's capabilities.  
But, if the wall represents a periodic function, the agent drives on a periodic path when the sensor has a narrow field of view. 
The agent's trajectory transitions to chaos when the sensor's field of view is increased.
Numerical experiments were performed with square, triangle, and sawtooth waves for the wall, to find this pattern. 
The bifurcations of the agents were analyzed, finding both border collision and period doubling bifurcations.

Detailed experimental results will be reported in the final version.
\end{abstract}

\section{Introduction}

In this paper we look at how a simple wall following system leads to chaos, when modeled as a mathematical system.  
This shows the importance of understanding chaos in the creation of artificial intelligences for robots. 
The model was originally described in the paper ``Chaos in Wall Following Robots" by Steve Mesburger and Alfred Hubler. A model for a simple wall following robot was described and it was shown that when following a sinusoidal wall the robot's behavior would undergo a transition to chaos via period doubling as the robot's field of view increased \cite{chaos-wall}.
In this paper we extend this research by walls with discontinuous corners and show that the robot still experiences a transition into chaos.

\section{Model}
The model for this simulation is small robot agent with an independent drive train and a simple forward pointing boundary sensor \cite{chaos-wall}.
An independent drive train \footnote{commonly found on tanks and the Mars Exploration Rovers} allows the vehicle to turn in place by driving the wheels on each side in opposite directions.
The boundary sensor is only able to minimum distance to the boundary and not the shape or direction of the boundary. 

Furthermore the sensor has field of view limited to $2 \Phi$. Where $\Phi$ is one of parameters varied in the experiment.

The most unrealistic part of this robot is that it can change its angular velocity (the rate at which it is turning)
instantaneously, allowing the robot to switch from a hard left turn to a hard right turn and back.
This can happen when the robot's sensor detects a wall at the edge of it's vision and then turns or moves away from the wall. 
This is not physically possible but as long as the speed is low this approximates how real vehicles behave.
For the purpose of modeling the robot also maintains a constant velocity which we convivially set to one.

As such the robots position can be represented by an $(x,y,\theta)$ vector where $\theta$ is the counter-clockwise angle of the robot's direction of travel from the positive x-axis. 
The velocity of the robot is represented by $v$ and the angular velocity is $\omega$. (this makes the last equations somewhat trivial)
The robot can the be represented as a series of deferential equations \cite{chaos-wall}.

\begin{align}
\frac{d x} {d t} &= v \cos \theta \\
\frac{d y} {d t} &= v \sin \theta \\
\frac{\delta \theta} {\delta t} &= \omega
\end{align}

Also the model dose not enforce the boundary by preventing the robot from crossing it.  
So it may be better to think of the boundary as the edge of a no entry zone for the robot instead of a physical barrier.

\begin{figure}[h]
\centerline{
\epsfig{figure=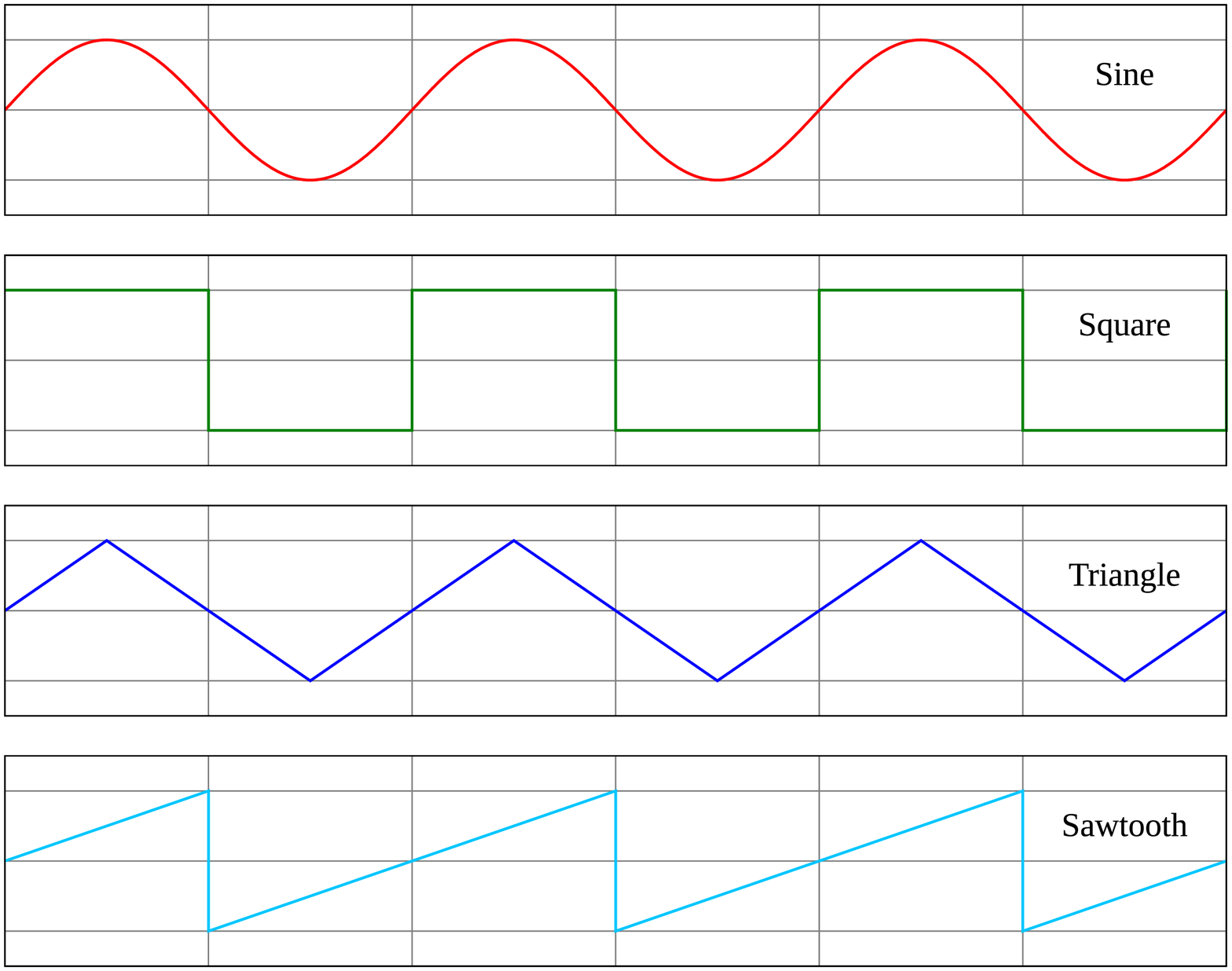,width={3.3in}} }
\caption{Four common types of waves.  The triangle, square, and sawtooth waves were the focus of this study.}
\end{figure}

The boundary is modeled as one of the waves.  
The wave travels along the X axis with an amplitude of one and a wavelength ($\lambda$) of 10. 
There waves are studied in this paper, the square wave, the triangle wave and the sawtooth wave. 
A sinusoidal wave has been studied previously \cite{chaos-wall}.

These waves can be represented using the following equations.
\begin{align}
 y_{sawtooth} (x)& = x - \lfloor x \rfloor \\
 y_{triangle} (x)& = \arcsin ( \sin ( x ) ) \\
 y_{square} (x)&= -1 ^ {\lfloor x \rfloor} 
\end{align}

However, for efficiency of the calculations, and because the boundaries the vertical disjoint parts of the wave are part of the boundaries each wave was represented as a series of points representing one period of the wave.
(Actually more than one period is represented to avoid issues with the robot around the boundaries.)
For example the square wave with $\lambda$ being the wavelength was represented by the following series of points.
$(-.5\lambda , -1 ) , ( 0 , -1 ) , ( 0, 1 ) , ( .5 \lambda , 1 ) , (.5 \lambda , -1 ) , ( \lambda , -1 ) $

\section{Procedures}
A model for the robot was created using Mathematica\texttrademark. 
Each step of the robot was calculated using Runge-Kutta 4 with a $\Delta t = 0.1$. 
Smaller values for $\Delta t$ did not improve the results significantly.

The largest calculation of the problem it to determine the sensor value, or the minimum distance to the boundary. 
For the each segment of the wall with the robot's range the closest point to the robot was calculated 
as well as any intersections with the edges of the robot's field of vision.
The closest of all these points was then used to calculate the sensor value.

$R$ represented the maximum range of the sensor. 
$r$ represented the range at which the robot would move straight or 
how far it tried to say away from the wall. 
A scaling value $\alpha$ of the angular velocity might also be appropriate.
Finally $\Phi$ was half of the robot's angular vision, it could range $(0,\pi]$.  
$R$ was taken to be 4, $r$ was taken to be 3, and $\alpha$ was taken to be $1.0$ . 

The results were then graphed on X-Y, X-$\Theta$, Y-$\Theta$.  With the X values were taken modulo $\lambda$ (10) so that they coincided with the period of the boundary.

\subsection{Calculation of Sensor Value}
In order calculate the closest point on the boundary to the robot agent with in it's sensor range several different distances were calculated and the minimum of them was used.
The first, trivially, distance is the maximum range of the sensor $R$.
Secondly the distance to each selected points on the boundary that have been tested to be in the robot's sensor range. 
These points are divided into two further groups, points on the intersection of the boundary and the edge of the robot's vision, and the closest point on each segment to the robot's location. 
It can be shown that these points will always contain the closest point to the robot within its vision.

The point on a segment closest to another point (the robot's location in this case) was found using parametric equations to reduce machine rounding errors and minimize the calculation time.
$\frac{ | (s_2 - s_1) \times ( p - s_1) | } { | s_2 - s_1 | ^ 2 } =  u$
Then:
\[
\text{Closest Point} =
\begin{cases} 
s_1 & \text{if } u \leq 0 \\
s_2 & \text{if } u \geq 1 \\
(s_2 - s_1) u + s_1 & \text if 0 < u < 1 
\end{cases}
\]

These points are then tested to make sure that they have the right angle from the robot to be seen by its sensors and are close enough to the robot to be seen.  Point that are not in the robots sensor range are removed. 
Each segment is then tested to see if any point on it intersects with either edge of the robot's sensor range.
The intersection of segments $(\hat{p_1},\hat{p_2})$ and  $(\hat{p_3},\hat{p_4})$ was then found by:\cite{algorithms}.

\begin{align}
u& =\frac{(x_4-x_3)(y_1-y_3)-(y_4-y_3)(x_1-x_3)} {(y_4-y_3)(x_2-x_1)-(x_4-x_3)(y_2-y_1)} \\
\text{intersection} & = \hat{p_1} + u ( \hat{p_2} - \hat{p_1} )
\end{align}

\section{Numerical Results}

The tree different boundary waves all followed the same pattern as $\Phi$ increased from $0$ to $\pi$ 
When $\Phi$ was close to zero the robot tends to hug the wall very closely.  In places where the is a long straight wall with no obstacles you can see the robot follow the wall all along it's path.  
The figures here were made by taking the point form each time step and plotting $(x \mod \lambda, y)$ where $\lambda = 10$ in this case.
\begin{figure}
\centerline{
\epsfig{figure=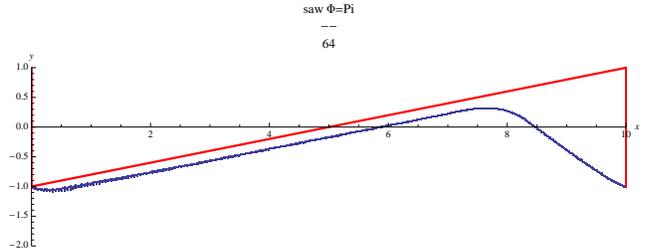,width={3.3in}}
}
\caption{ Plot of the robot following a sawtooth wave, when $\Phi=1/64 \pi$. Notice how the robot is willing to follow the straight segment as long as it goes.} 
\end{figure}
You also see the robot tends to head directly for corners, just narrowly passing them and then quickly stabilizing on the segment.
Furthermore as $R$ remains constant decreasing $r$ causes the robot to follow the wall more closely. 
However when $r=R$ the robot no longer turns counter clockwise. 
Similarly when $r=0$ the robot will only turn counter clockwise.
Both of these cases are less interesting as the robot tends to either go in circles or drive off in a straight line.
\begin{figure}
\centerline{
\epsfig{figure=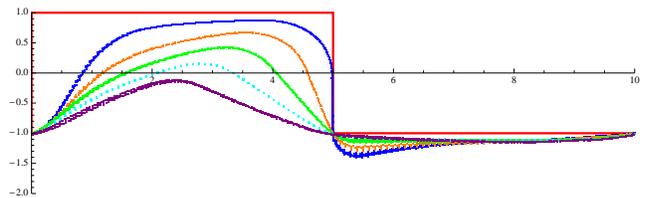,width={3.3in}}
}
\caption{ Plots of the xy movement when the sensor range $\Phi=1/64 \pi$, $R=4$ and $r$ in the range $[1.5,3.5]$.}
\end{figure}

As the value of $\Phi$ increases robot's path becomes slightly chaotic.
While the robot's path remains in a very narrow, and stable channel within that channel the path is chaotic. 
With the sawtooth wave this behavior becomes most pronounced when $\Phi \approx 19 \pi / 64$ .  
 
\begin{figure}
\centerline{
\epsfig{figure=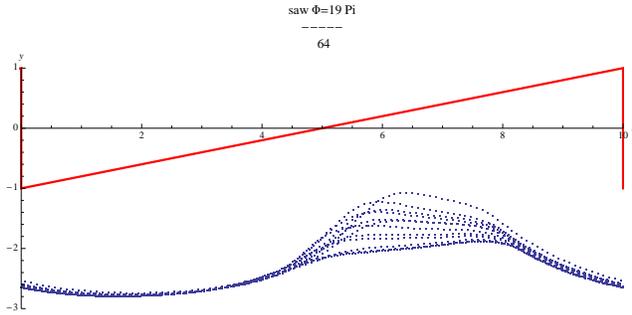,width={3.3in}}
}
\caption{ Path of robot against the sawtooth wave show in experience chaos even at low values of $\Phi$.}
\end{figure}
 
\begin{figure}
\centerline{
\epsfig{figure=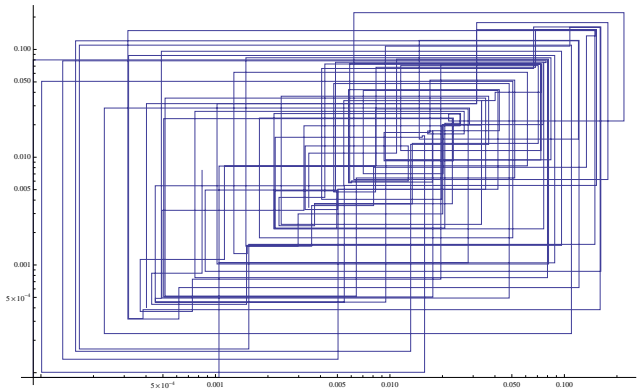,width={3.3in}}
}
\caption{ Cobweb diagram of the robot's $y$ position when as it passes a point where $x \mod \lambda = 0$.
The plot show $ \log ( y + 2.65 )$ as this makes the  lower left hand section of the graph easier to read.}
\end{figure}

Verhulst diagrams or more commonly cobweb plots were created of a single path to better view robot's behavior.
To do this a point from each period of the robot's path was selected.  
For simplicity the points were when the x-coordinate was a multiple of the wavelength of the boundary.  
Then either the $y$ or $\theta$ values could be used to make the Verhulst diagram.  
The $y$ value was chosen arbitrary. 

In fact it was found that the robot tends to exhibit this chaotic behavior even at low $\Phi$ where it appears stable.
This may be a property of having sharp corners in the boundary wave that are not present in a sinusoidal wave.

\begin{figure}
\centerline{
\epsfig{figure=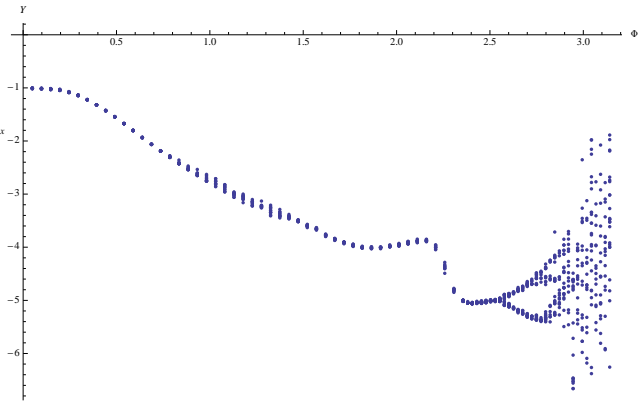,width={3.3in}}
}
\caption{ A Bifurcation diagram of sawtooth wall showing a transition to chaos through period doubling}
\end{figure}

For all three waves the chaotic patterns appeared to die down around $\Phi \approx .5 \pi  $. 
Then after that all three waveforms undergo a transition to chaos.
To view the transition bifurcation diagrams were created for each type of wall.  
Like the Verhulst diagrams the value of $y$ at the when $x$ is a multiple of $\lambda (10)$ were taken from each simulation of the robot.
The first 30\% of the values were discarded to ensure that the system and converged to it's attractor.

The sawtooth boundary produced an period doubling bifurcation diagram.  
Where the full transition to chaos occurred when $\Phi \approx 121 / 128 \pi $
However an exception was also found when $\Phi = 15 / 16 \pi $.
The robot moved in the negative direction by turning  counter clockwise in ellipses.
It may be that this is another attractor for the sawtooth boundary regardless of the $\Phi$ value.

\begin{figure}
\centerline{
\epsfig{figure=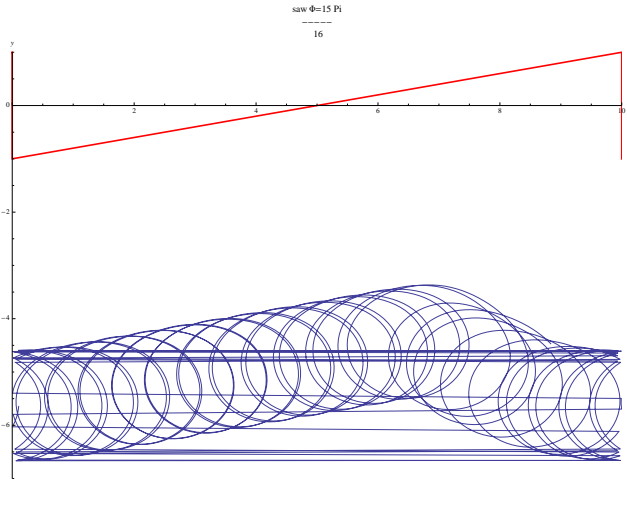,width={3.3in}}
}
\caption{ A plot of the robot's movement backwards along a sawtooth boundary.}
\end{figure}

When the robot was following a triangle wave it was found that the $\theta$ values provided a better bifurcation diagram than the $y$ values.
The triangle wave boundary first cause chaotic behavior around $\Phi \approx 3 / 16 \pi$
The sharp disconnect when $\Phi \approx 3 / 4 \pi$ in the bifurcation diagram appears to be an indicator of a second attractor. 
\begin{figure}
\centerline{
\epsfig{figure=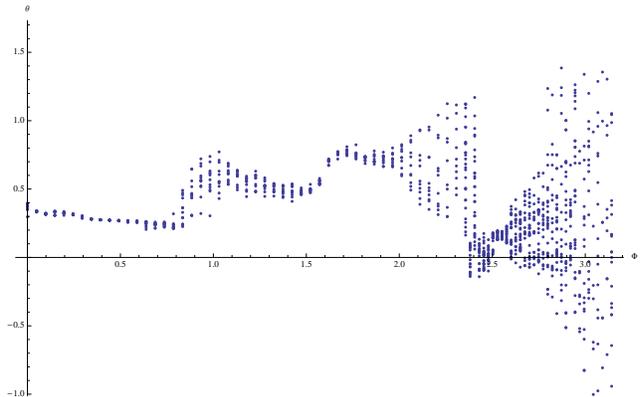,width={3.3in}}
}
\caption{ A Bifurcation diagram of triangle wall }
\end{figure}

The square boundary gave results looking more similar to that of the triangle wave then the sawtooth wave.
It's first expression of chaos occurs around $\Phi \approx \pi / 4 $ 
Also around  $\Phi \approx 49 / 64 \pi $ it too has a disconnect, with initial starting values of $( 0, -2 ,0) $.
In order to make the bifurcation diagram simulations were run for each $\Phi$ value with a starting points at $(0, y, 0) $ where for every half unit between $-1$ and $-5$. 
These two strange attractors actually curl around each other when visualized in 3d.
\begin{figure}
\centerline{
\epsfig{figure=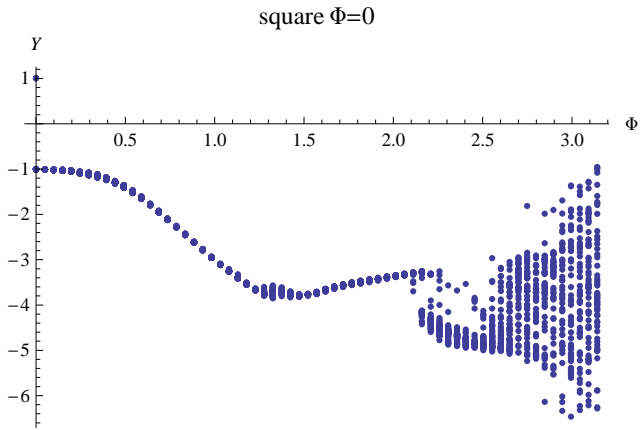,width={3.3in}}
}
\caption{ A Bifurcation diagram of Square wall showing a very sharp transition around $\Phi = \frac{3 \pi}{4} \approx 2.3$ }
\end{figure}

\begin{figure}[h]
\centerline{
\epsfig{figure=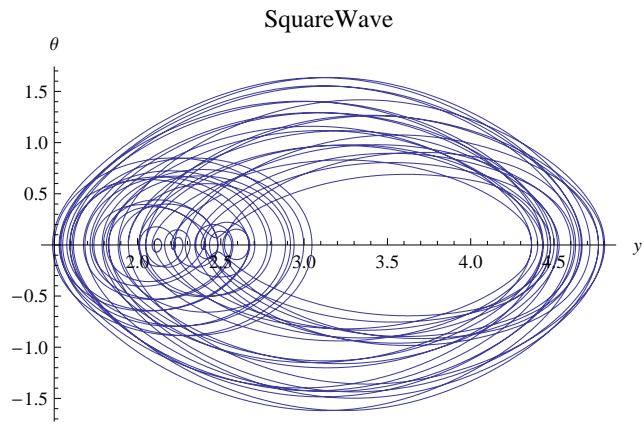,width={3.3in}}
}
\caption{A Poincare Plot of the y and $ \theta $ values when x is multiple of 10.  }
\end{figure}

\section{Conclusion}	\label{S:conc}
In this work it was show that the behavior of a simple control system can under go a chaotic transition from a change in the environment. 
A consistent pattern was still observable through out these changes.

Further improvements to the algorithm could be made by creating a spatial hash table of segments that maps the location of the robots to a list of segments that are with in the range of the robot. 
 This would reduce the number of tests the simulation performs by as much as 50\%.

\section{Future Work}

Further work beyond optimizations of the algorithm involve adding additional robot agents for make a swarm. 
The swarm of wall following robots would interact with each other and the wall, which leads to a rich dynamical system. 
Research on the conditions when the swarm behaved chaotically and when it did not would be done.

\begin{itemize}
\item Build a swarm of wall following robots
\item Experiment with different cognitive behaviors in the agents
\item Use agents with differing abilities in both sensors and cognition
\item Combine real and virtual robot agents in a simulation
\end{itemize}

\bibliography{wall}
\end{document}